\documentclass[epj,dvips]{svjour}
\usepackage{latexsym}
\usepackage{graphics}

\usepackage{amsmath,amssymb} %% 20161119 added
\usepackage{widetext} %% 20161119 added

\usepackage{color} %% 20170129 added

\begin{document}
\title{Theoretical and numerical analysis of a heat pump model utilizing
Dufour effect}
%\author{First author\inst{1} \and Second author\inst{2}% etc
\author{Minoru Hoshina and Koji Okuda
% \thanks{\emph{Present address:} Insert the address here if needed}%
%
}                     % Do not remove
%\institute{Insert the first address here \and the second here}
\institute{Division of Physics, Hokkaido University, Sapporo 060-0810, Japan}
\date{Received: date / Revised version: date}
\abstract{
  A
  heat pump model utilizing the
  Dufour effect is proposed, and studied 
 by numerical and theoretical analysis.
 Numerically, we perform MD simulations of this system 
 and measure the cooling power and the coefficient of performance (COP)
 as figures of merit.
 Theoretically, we calculate the cooling power and the COP from 
 the phenomenological equations describing this system 
 by using the linear irreversible thermodynamics and 
 compare the theoretical results with the MD results.
%
% \PACS{
%       {PACS-key}{discribing text of that key}   \and
%       {PACS-key}{discribing text of that key}
%      } % end of PACS codes
} %end of abstract
\maketitle
%
%\section{Introduction}
%\label{intro}
\section{\label{sec:intro}Introduction}
The Dufour effect \cite{LDufour} induces the temperature difference from the mole fraction
difference in the mixture fluid system %multi-component system
as the Peltier effect \cite{Peltier} does from the electric potential difference.
Although the Peltier effect is widely applied
to various heat pumps \cite{Rowe,Goldsmid},
it has not been studied whether an application of the Dufour effect to heat pumps
is possible or not.
In this paper, we propose a heat pump model utilizing the
Dufour effect and study this model by numerical and theoretical analysis.

The Dufour effect is well studied by the experiments
\cite{LDufour,Miller,Horne1978,Horne1980,Waldmann1947a,Waldmann1949,Korzhuev,KEGrew}
and theoretical approaches such as the linear irreversible
thermodynamics \cite{Degroot,Horne1973}, the Chapman-Enskog theory
\cite{SChapman,Chapman1}, the
phenomenology \cite{Waldmann1943a}, and other methods
\cite{Streather2000,RGMortimer1980}.
In 1873, L. Dufour mixed the two gases of different
molecular-weights and discovered a temperature fall in the
higher-molecular-weight gas during the diffusive mixing process \cite{LDufour}.
The theory describing this effect was first developed by Chapman and
Enskog by
applying the kinetic theory to the microscopic analysis of
the binary gas mixture \cite{SChapman}, in
which the temperature $T$ and the number-densities of molecules
 $n_A$ and $n_B$ of the two components $A$ and $B$ are non-uniform in space.
They derived that the heat current ${\vec J}_Q$ can be written as
\begin{align}
  {\vec J}_Q =-\kappa {\vec \nabla}T
  -nk_BT^2D''{\vec \nabla}x_A,\label{122511_5Dec13}
\end{align}
where $D''$ is the Dufour coefficient,
$\kappa$ is the thermal conduction coefficient,
$k_B$ is the Boltzmann
constant, $n$ is the total number-density of all the components,
i.e. $n=n_A+n_B$, and $x_A$ is 
the mole fraction of the component $A$ defined as
$x_A \equiv n_A/n$.
The result $D''<0$ can also be derived from their theory when the
molecular mass of the component 
$A$ is lower than that of the component $B$ (i.e. $m_A<m_B$)
in some special cases of the intermolecular potential.
This result implies that the heat current tends to flow from $A$-rich
part to $B$-rich part, which is consistent with the above experiment by Dufour.

The organization of this paper is as follows.
We construct a heat pump model utilizing the
Dufour effect in Section~\ref{sec:model}, and the usefulness of this model
as a heat pump
is confirmed numerically using the molecular
dynamics (MD) simulation \cite{Allen} in Section~\ref{sec:md1}.
Next, by using the linear irreversible thermodynamics \cite{Degroot},
we theoretically analyze this model in a simple case
where the heat pump is driven very slowly and attached to the two heat baths
whose temperature difference is zero or small,
and compare the theoretical results with the data obtained numerically by
the MD simulation in Section~\ref{sec:theroy}.
Finally, we summarize this study in Section~\ref{sec:summary}.

\section{\label{sec:model}Model}
 The main idea of our model is the following.
Since the Dufour effect occurs only
during the transient diffusive mixing process,
as far as we know from the previous experimental studies
\cite{LDufour,Miller,Horne1978,Horne1980,Waldmann1947a,Waldmann1949,Korzhuev,KEGrew},
it is difficult to keep the Dufour effect constant like the steady state
of the Peltier effect.
For this reason, we need a process
that separates the
components of the mixture, besides the diffusive mixing process.
In our model, an external electric field  is used for the
separation of the mixture.

\begin{figure}
\resizebox{0.5\textwidth}{!}{%
  \includegraphics{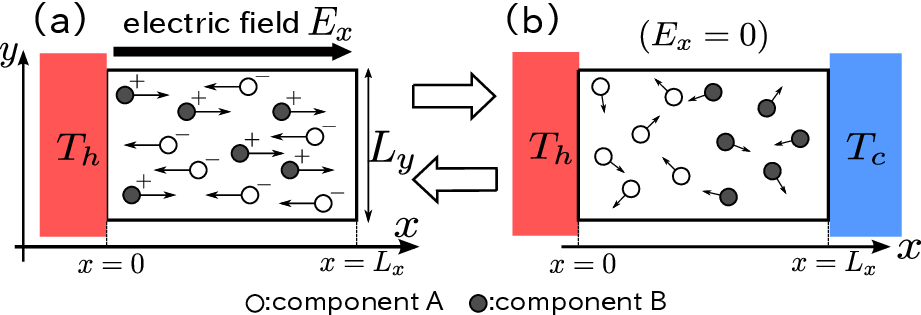}
}
\caption{\label{fig2}
 Schematic illustration of the system and the processes of the heat pump
 model.
 (a) In the {\itshape separating process}, only the heat bath with $T_h$ is
 attached and the electric field $E_x$ is applied.
 (b) In the {\itshape mixing process}, both of the heat baths with $T_h$
 and $T_c$ are attached and the electric field is turned off.}
\end{figure}

Consider a gas mixture of the two components $A$ and $B$,
and
assume that the molecular mass of the component $A$ is lower than
that of $B$,
so that $m_A < m_B$.
To separate the mixed components into $A$ and $B$ by an electric field,
electric charges $q_A$ and $q_B$ are given to each molecule of $A$ and
$B$, respectively, and we assume
$q_A = -q$ and $q_B=q \,(q>0)$ for simplicity.
Though the molecules have electric charge, Coulomb interaction between
them is ignored throughout the paper
\footnote{We note that our purpose in this paper is to suggest the possibility of the heat pump utilizing the Dufour effect.
Though we use the electric field and charged particles to separate the components clearly, we consider that this method can be replaced with another such method as using gravity to realize this heat pump.
This is discussed as a remaining problem in the last paragraph of Section~\ref{sec:summary}.}.
The particle numbers of the components in the system are $N_A$ and
$N_B$, and
other properties of the components $A$ and $B$
such as the particle interaction or the shape of the molecules
are assumed to be identical.

This gas mixture is contained in the system as schematically depicted in
Figure~\ref{fig2}.
The system is a two-dimensional rectangle with the size
$L_x \times L_y$.
To pump a heat from the heat bath with a low temperature $T_c$ to the
heat bath with a high temperature $T_h$, two procedures, (a)
{\itshape separating process} and (b) {\itshape mixing process}
are alternately repeated.
The details of these processes depicted in Figure~\ref{fig2} are described
as follows.
\begin{description}
 \item[(a) Separating process :] during this process, the heat bath with
	    $T_h$ is attached to the boundary at $x=0$, and 
	    the insulated wall is placed on the boundary at $x=L_x$.
	    Furthermore, a static external electric field $E_x=E(>0)$ is
	    applied in the $x$-direction.
	    After continuing this process for a duration $\Delta t_{\text{sep}}$, the
	    system is switched to the mixing process.
 \item[(b) Mixing process :]  during this process, the heat baths with
	    $T_h$ and $T_c$ $(T_h>T_c)$ are attached to the boundary at
	    $x=0$ and $x=L_x$, respectively, and the electric field
	    is turned off ($E_x=0$).
	    After continuing this process for a duration $\Delta t_{\text{mix}}$, the
	    system is switched to the separating process.
\end{description}
In the separating process, the components of the gas mixture are
separated by the external field $E_x$ so that 
a negative gradient of the mole fraction
$\partial x_A/\partial x<0$ is established.
The heat energy due to the work done by the external field $E_x$ leaks into
the heat bath
with the temperature $T_h$, and the system approaches the
equilibrium state of the total system at the temperature $T_h$
if the duration
$\Delta t_{\text{sep}}$ is taken sufficiently long.
In the mixing process, a diffusive mixing of the components $A$ and $B$
occurs.
As seen from equation~(\ref{122511_5Dec13}) and $m_A<m_B$ (therefore $D''<0$),
a heat current flows in the negative $x$-direction due to the
Dufour effect so that an amount of heat is expected to be pumped from
the heat bath with $T_c$ to the heat bath with $T_h$.

\section{\label{sec:md1}MD simulation of the model}
\subsection{\label{sec:md1-model}The simulation model}
In our simulation model, the time evolution of the system is governed by a
Hamiltonian 
\begin{align}
  \mathcal{H}&=\displaystyle \sum_{i=1}^{N} \frac{\vec{p}_i^2}{2m_i}
  + \sum_{i<j} U^{\text{int}}(|\vec{r}_i - \vec{r}_j|)
  - \sum_{i=1}^{N} q_i E_x(t) \tilde{x}_i, \nonumber\\
 &\qquad\qquad\qquad\qquad\qquad\qquad\quad
 (N \equiv N_A+N_B),\label{153558_12Dec13}
\end{align}
where ${\vec p}_i,{\vec r}_i,m_i,q_i$, and $\tilde{x}_i$ denote the momentum,
position, mass, electric charge, and $x$-coordinate of the
$i$th particle, respectively.
$U^{\text{int}}$ denoting the interaction potential for the center-to-center
distance $r$ of the particles is taken to be a hard Herzian
potential \cite{Love,Yukawa,Yuge},
\begin{align}
  U^{\text{int}}(r)= \begin{cases}
Y|\sigma - r|^{\frac{5}{2}}& (r \leq \sigma)\\
     0  &  (\sigma < r)
		    \end{cases},\label{163720_14Dec13}
\end{align}
where $\sigma$ is the diameter of the particle, and a constant $Y$ is taken to be
$Y=10^5 \epsilon \sigma^{-\frac{5}{2}}$ with an energy unit $\epsilon$.
The external electric field $E_x(t)$ is defined as
\begin{align}
 E_x(t) =
  \begin{cases}
   E & (\text{in the separating process})\\
   0 & (\text{in the mixing process}),
   \end{cases}
\end{align}
where $E$ is a positive constant.
\textcolor{black}{Note that the electric charge of particles is used only to separate
the components and for simplicity Coulomb interaction between them is
ignored in our simulations.}

The periodic boundary condition is imposed in the $y$-direction.
The boundary of the $x$-direction at $x=L_x$ is
the elastically reflecting wall in the separating process,
and the thermalizing wall \cite{Tehver} with the 
temperature $T_c$ in the mixing process.
The boundary at $x=0$ is also the thermalizing wall with the
temperature $T_h$ in both of the processes.
If a particle with the mass $m$ collides with the thermalizing
wall with the temperature
$T$, its velocity is stochastically changed to a value
${\vec v}=(v_x,v_y)$ according to
the distribution functions
\begin{align}
     P_x(v_x)&=\frac{m}{k_BT}|v_x| \exp \Big(-\frac{mv_x^2}{2k_BT}\Big),\nonumber\\
 &\qquad\quad \text{where}
 \begin{cases}
   v_x>0 ~\text{at}~x=0\\
   v_x<0 ~\text{at}~x=L_x,
 \end{cases}\\
    P_y(v_y)&=\sqrt{\frac{m}{2\pi k_BT}} \exp \Big(-\frac{mv_y^2}{2k_BT}\Big),
\end{align}
which ensure that the temperature of the equilibrium system becomes $T$.

In the following simulations,
we use the scale units as
$m_A\equiv 1$,$\sigma\equiv 1$,$\epsilon \equiv 1$,
$q\equiv 1$,
and $k_B\equiv 1$, which define the units of
mass, length,
energy, electric charge, and temperature, respectively.
The time evolution of the system is performed by
the velocity-Verlet scheme \cite{Allen} with the time step
$\delta t=0.0005$.
\subsection{\label{sec:md1-result}Results of the simulation}
\begin{figure}
\resizebox{0.48\textwidth}{!}{%
  \includegraphics{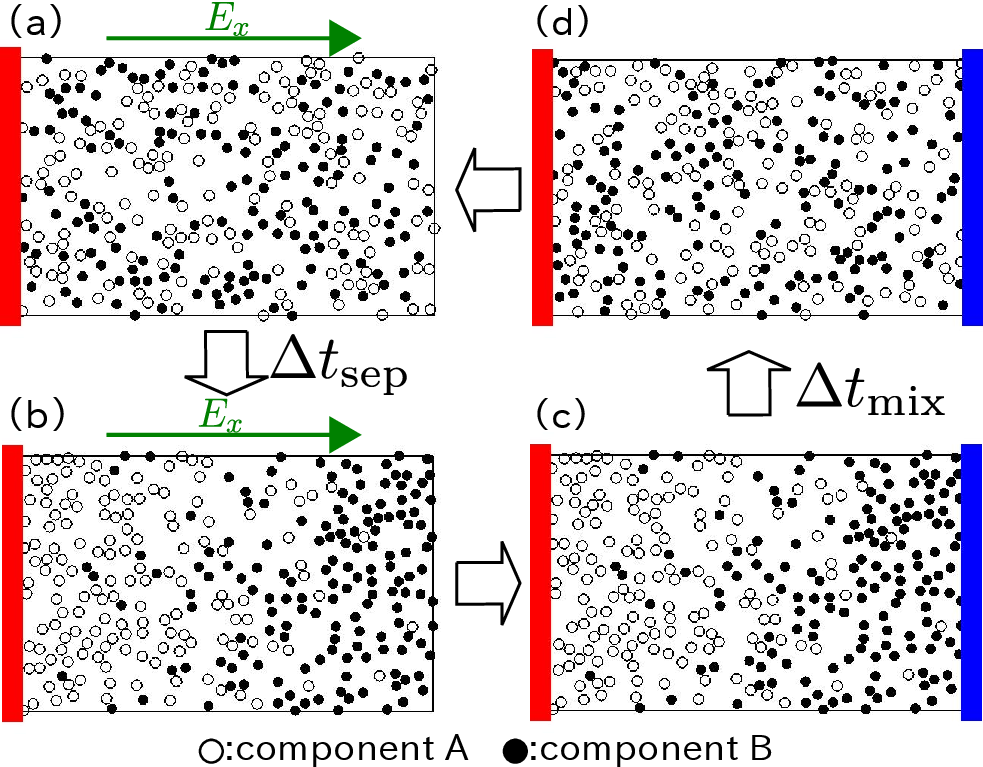}
}
\caption{\label{fig4}
 Example of the snapshots of the system.
 (a)\,The beginning of the separating process.
 (b)\,The end of the separating process.
 (c)\,The beginning of the mixing process.
 (d)\,The end of the mixing process.
 The white disks and the black disks indicate the
 low-molecular-weight
 component $A$ and the high-molecular-weight component $B$, respectively.
 }
\end{figure}

Figure~\ref{fig4} shows an example of the snapshots of the system.
In the simulation, the system size is $L_x\times L_y=40\times 25$, the numbers
of the particles are $N_A=N_B=150$,
the external field is $E=0.1$, 
and the temperatures of the heat baths are $T_h=1.01$ and $T_c=0.99$.
Each particle of the components $A$ and $B$ has the mass $m_A=1$ and
$m_B=10$, and the electric charge $q_A=-1$ and $q_B=1$, respectively.

\begin{figure}
\resizebox{0.48\textwidth}{!}{%
  \includegraphics{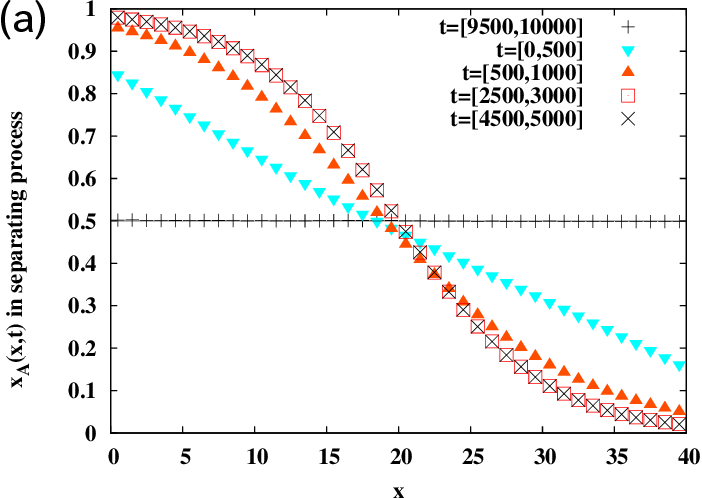}
}
\resizebox{0.48\textwidth}{!}{%
  \includegraphics{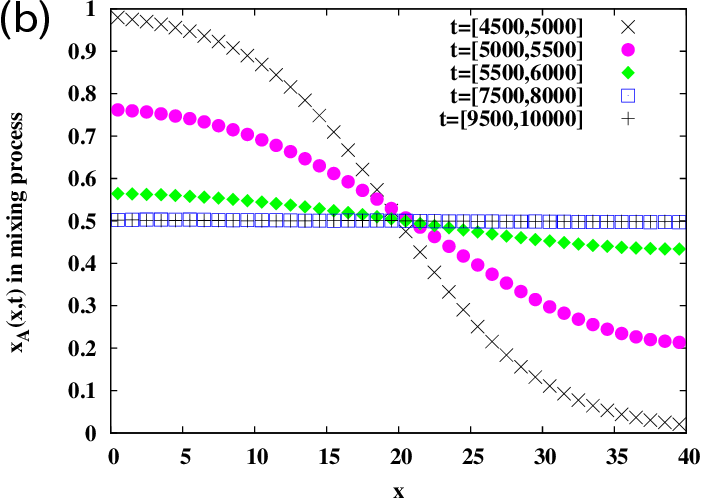}
}
\caption{\label{2dmd_xAloc-t} 
 The mole fraction profile $x_A(x,t)$ in the separating process
 ($0 < t \leq 5000$), and in the mixing process
 ($5000 < t \leq 10000$), with
 $\Delta t_{\text{sep}}=\Delta t_{\text{mix}}=5000$.
 A curve of $t=[t_1:t_2]$ means a profile
 averaged over the time between $t_1 \leq t \leq t_2$. 
 Furthermore, the MD data were averaged over 2640 cycles.
 }
\end{figure}

From these snapshots, we can confirm that the components $A$ and $B$ are
separated by applying the external field $E$ in the separating process
and the components are diffusively mixed when the external field is turned
off in the mixing process.
This result can quantitatively be verified in Figure~\ref{2dmd_xAloc-t} 
which shows an example of the time evolution of the mole fraction
profiles $x_A(x,t)$
in the mixing process and the separating process,
where we calculated $x_A$ by dividing the system
into 40 subsystems in the x-direciton.

\begin{figure}
\resizebox{0.48\textwidth}{!}{%
  \includegraphics{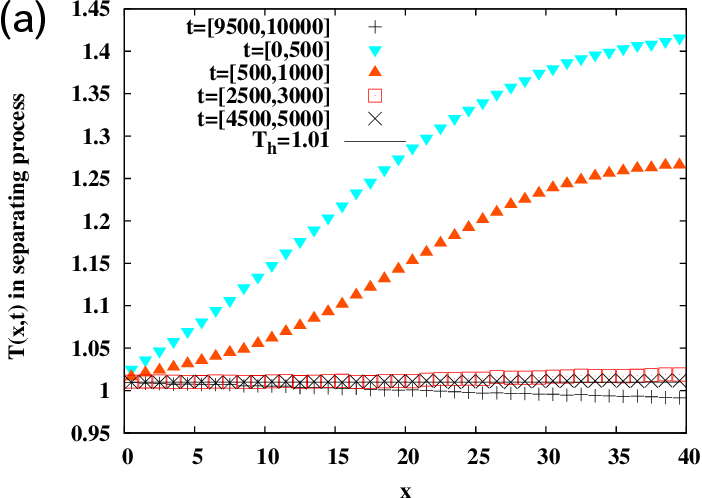}
}
\resizebox{0.48\textwidth}{!}{%
  \includegraphics{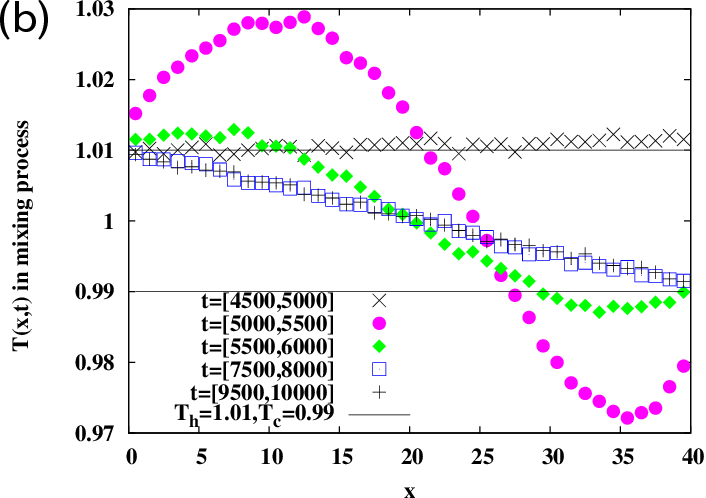}
}
\caption{\label{2dmd_Tloc-t}
 The temperature profiles $T(x,t)$ in the separating process
 ($0 < t \leq 5000$), and in the mixing process
 ($5000 < t \leq 10000$).
  A curve of $t=[t_1:t_2]$ is drawn following the same rule with
 Figure~\ref{2dmd_xAloc-t}. 
 The solid lines denote the heat bath temperatures $T_h=1.01$ and $T_c=0.99$.
 }
\end{figure}

Figure~\ref{2dmd_Tloc-t} depicts typical results of the time evolution
of the temperature profiles $T(x,t)$ of the system,
which is calculated from the kinetic energy as
$T=\frac{1}{2N_x} \sum_{j=1}^{N_x} m_j {\vec v}_j^2$,
by using the same subsystems as in Figure~\ref{2dmd_xAloc-t} , where
 $N_x$ is the number of particles in
 the subsystem at position $x$, and $m_j$ and \textcolor{black}{${\vec v}_j$} are the mass and velocity
 of the $j$th particle in that subsystem,
  respectively.

\begin{figure}
\resizebox{0.48\textwidth}{!}{%
  \includegraphics{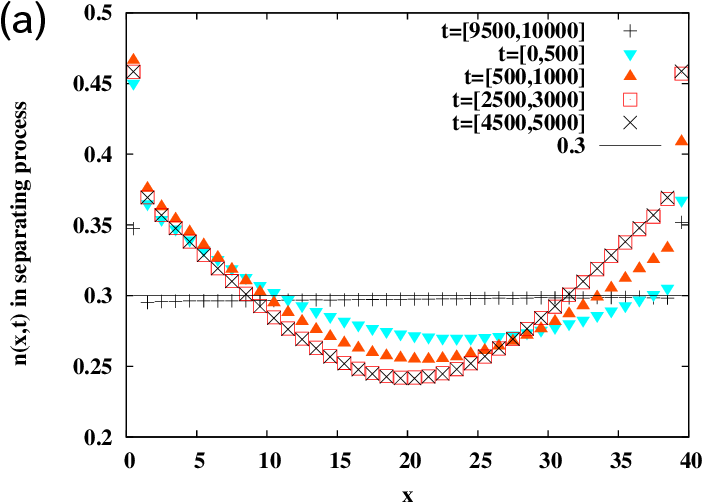}
}
\resizebox{0.48\textwidth}{!}{%
  \includegraphics{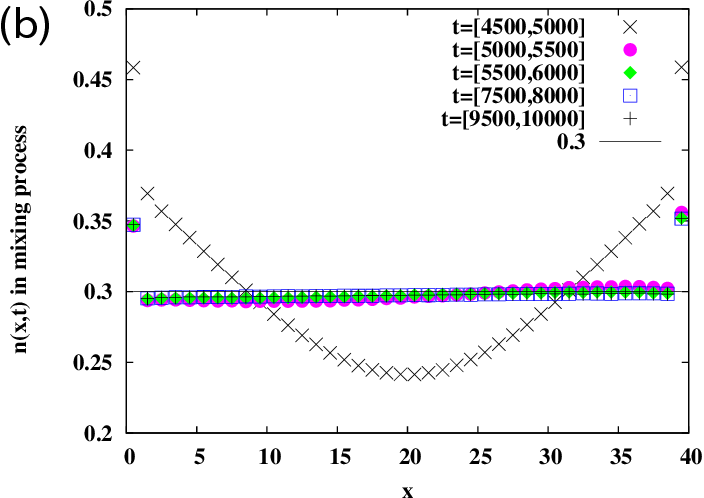}
}
\caption{\label{2dmd_Nloc-t}
 The number-density profiles of the particles $n(x,t)$ 
 in the separating process
 ($0 < t \leq 5000$), and in the mixing process
 ($5000 < t \leq 10000$).
  A curve of $t=[t_1:t_2]$ is drawn following the same rule with
 Figure~\ref{2dmd_xAloc-t}.
 The solid line denotes the average number-density $\overline{n}=0.3$.
 }
\end{figure}
\begin{figure}[tb]
  \resizebox{0.48\textwidth}{!}{%
    \includegraphics{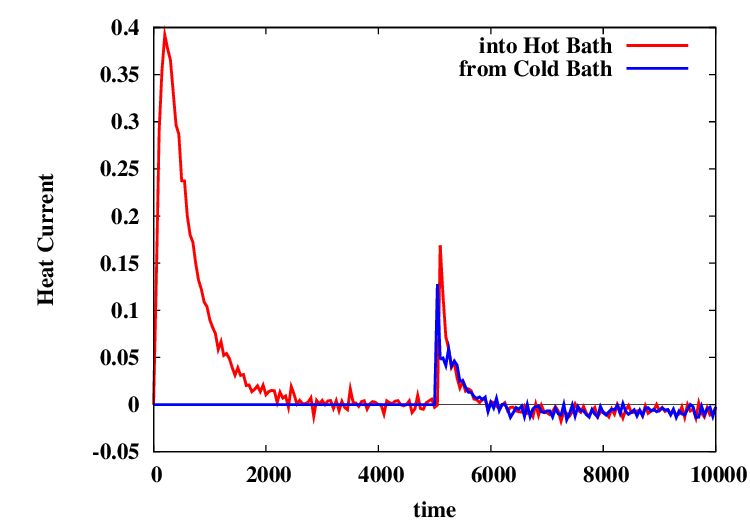}
}
\caption{\label{2dmd_Jch-t}
 Time dependence of the heat currents $\dot{Q}_c(t)$
 flowing from the heat bath with $T_c$, and $\dot{Q}_h(t)$
 flowing into 
 the heat bath with $T_h$. The system is in separating process when
 $0 < t \leq 5000$, and in the mixing process when
 $5000 < t \leq 10000$.
 The MD data were averaged over 2640 cycles.
 }
\end{figure}
In the separating process, after the sudden increase of the temperature
due to the heat produced from the work done by the external field 
$E_x$, the heat in the system gradually leaks into the heat bath with
$T_h$, and then the
total system reaches the equilibrium state at the temperature $T_h$.
In contrast, in the mixing process, the system reaches a nonequilibrium
steady state of heat conduction with a spatially linear temperature
profile. 
We note that the temperature profile of the data t=[5000,5500] in Figure~\ref{2dmd_Tloc-t}b
can be explained as follows.
At the early stage of the mixing process, in the middle region of the
system, the Dufour effect due to the large mole-fraction gradient
(see the data t=[4500,5000] in Fig.~\ref{2dmd_xAloc-t}b) causes a large heat flow in
the negative $x$-direction, while in the regions of both ends, the heat flow
by the Dufour effect is small due to the small mole-fraction gradient.
Accordingly, the temperature profile develops a maximum and a minimum
near both ends, as shown by the data t=[5000,5500] in Figure~\ref{2dmd_Tloc-t}b.
Once the maximum and the minimum of the temperature profile are formed,
the heat pump becomes functional by the heat flow in the negative
$x$-direction due to the temperature gradient near both ends
at this early stage of the mixing process.

Figure~\ref{2dmd_Nloc-t} shows the time evolution of 
the number-density profiles of the
particles $n(x,t)$.
The peaks of $n(x,t)$ near the boundaries
seem to be essentially the same phenomena as the particle adsorption
at a hard wall reported in \textcolor{black}{references}~\cite{Henderson1983,Snook1978}.
We can find from Figure~\ref{2dmd_Nloc-t} that the profile $n(x,t)$
 in the mixing process instantly reaches the steady profile
 compared with the mole fraction $x_A(x,t)$ in Figure~\ref{2dmd_xAloc-t}
 and the temperature $T(x,t)$ in Figure~\ref{2dmd_Tloc-t}.
This result is assumed to hold in general
for the theoretical analysis in Section~\ref{sec:theroy}.

In Figure~\ref{2dmd_Jch-t}, we measured the heat currents
$\dot{Q}_h(t)$ flowing from the system into the heat bath
with $T_h$ and 
$\dot{Q}_c(t)$ flowing from the heat bath with $T_c$ into the
system.
Here, we calculated $\dot Q_\alpha$ by accumulating over the unit time
   the kinetic energy change ${m \over 2}(v_0^2 - v^2)$ at a particle collision
   with the \textcolor{black}{thermalizing wall} with $T_\alpha$ ($\alpha=h,c$), where $m$ is the mass of
   the particle and $v_0$ and $v$ are the velocities of the particle before and
   after the collision, respectively.
We can see that $\dot{Q}_h(t)$ has a peak
corresponding to the heat injection due to the external field in the
separating process, and the thermal
equilibrium state of the total system is realized at last.
The peaks of $\dot{Q}_h(t)$ and $\dot{Q}_c(t)$ in the
mixing process which have a similar profile imply that the heat flows
from the cold heat bath with $T_c$ toward the hot heat bath with $T_h$ through
the system.
Therefore we can see that a heat pump due to the Dufour effect is realized.

\begin{figure}
\resizebox{0.45\textwidth}{!}{%
  \includegraphics{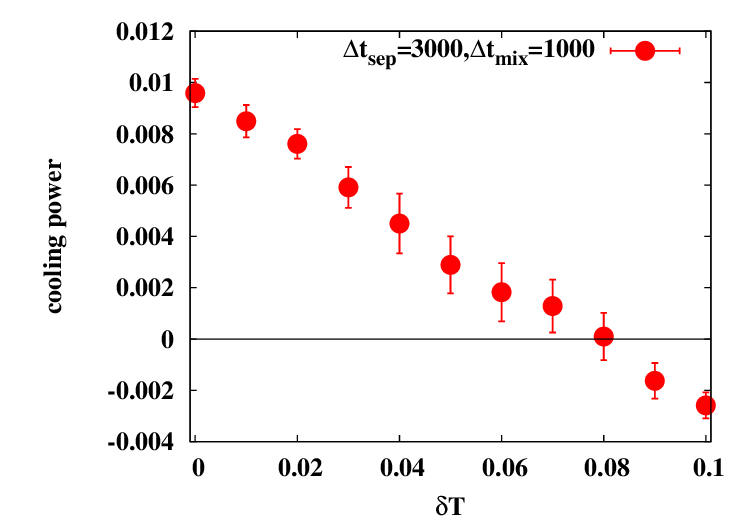}
}
\resizebox{0.45\textwidth}{!}{%
  \includegraphics{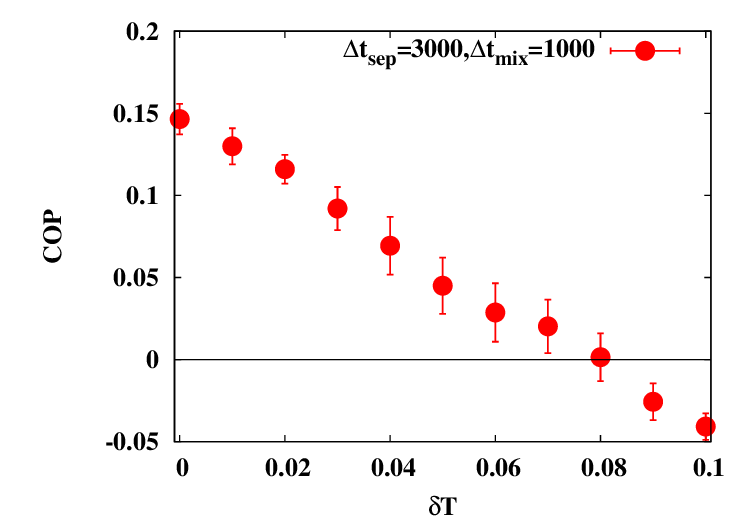}
}
\caption{\label{2dmd_COPcp-DelT}
 The temperature difference $\delta T=T_h-T_c$ dependence of the
 cooling power and the COP, with the process durations
 $\Delta t_{\text{sep}}=3000,\Delta t_{\text{mix}}=1000$,
 and the temperatures $T_h=1+\delta T/2$ and $T_c=1-\delta T/2$.
 The MD data were averaged over
345 cycles.
 }
\end{figure}

To confirm that our model is surely useful as a heat pump, we measured
the cooling power $\overline{\dot{Q}}_c$ and the coefficient
of performance (COP) $\epsilon$ defined as
\begin{align}
  \overline{\dot{Q}}_{\alpha}
  \equiv \frac{1}{\tau_1-\tau_0}
  \int^{\tau_1}_{\tau_0} \dot{Q}_{\alpha}(t)
  \,dt
  \quad (\alpha=h,c)  ,\label{123022_11Dec13} \\ 
 \epsilon \equiv \frac{\overline{\dot{Q}}_c}{\overline{\dot{W}}},
 \qquad\qquad\qquad\qquad \label{003700_24Nov11}
\end{align}
where $\tau_0$ is the relaxation time for the system to exhibit a steady
cyclic state, and
$\tau_1$ is chosen so that $\tau_1-\tau_0$ is an integer multiple of 
the cycle period $\Delta t_{\text{mix}}+\Delta t_{\text{sep}}$. 
$\overline{\dot{W}}$ in equation~(\ref{003700_24Nov11}) denotes the
average power
done by the external field \textcolor{black}{$E_x(t)$} per unit time,
which is calculated using the relation 
   $\overline{\dot{W}}=\overline{\dot{Q}}_h-\overline{\dot{Q}}_c$.
The cooling power  $\overline{\dot{Q}}_c$ and the COP  $\epsilon$
should be positive for a useful heat pump.
Figure~\ref{2dmd_COPcp-DelT} shows the $\delta T$-dependence of
$\overline{\dot{Q}}_c$ and $\epsilon$, where $\delta T\equiv T_h-T_c$.
  While the cooling power and the COP are surely positive when $\delta T$ is small,
  they become negative when $\delta T$ is large,
  because the heat pumping by the Dufour
  effect cannot overcome the temperature gradient between the heat baths.
Consequently, this numerical result implies that our model is useful as a heat pump when
the temperature difference $\delta T$ is sufficiently small and
$\Delta t_{\text{sep}}$ and $\Delta t_{\text{mix}}$ are suitably chosen.

\section{\label{sec:theroy}Theoretical analysis}
\subsection{\label{sec:COPcp_1}Expressions for the Cooling Power and the COP}
First, we consider a simple case that the heat baths have the same
temperature $T_0(=T_h=T_c)$, and assume that a process is switched to
another process after the equilibrium state is realized,
which means $\Delta t_{\text{sep}} \gg \tau_{\text{sep}}$
and $\Delta t_{\text{mix}} \gg \tau_{\text{mix}}$
where $\tau_{\text{sep}}$ and $\tau_{\text{mix}}$ are the relaxation
times of the system to 
the steady states in the separating process and the
mixing process, respectively.
To obtain simple expressions for the cooling power and the COP,
we assume that the mechanical equilibrium state
(see Chap.V-2 in Ref.~\cite{Degroot}) is
instantly realized in the mixing process.
This assumption means that the system satisfies
${\vec \nabla}p=n_A{\vec F}_A+n_B{\vec F}_B$
where $p$ is the pressure and ${\vec F}_k$ is the external force on each
particle of component $k$.
Therefore, the pressure gradient ${\vec \nabla}p$ vanishes as
\begin{equation}
 {\vec \nabla}p=0,\label{165229_5Mar14}
\end{equation}
in the mixing process where the electric field is turned off.
Furthermore, we assume that the number-density profile of the particles
$n(x,t)$ in the mixing process reaches the steady
profile instantly compared with the mole
fraction profile $x_A(x,t)$ and the temperature profile $T(x,t)$,
which is confirmed to hold in our system from the
numerical results in Figures~\ref{2dmd_xAloc-t}-\ref{2dmd_Nloc-t}.
From this assumption
$n(x,t)$ is approximately regarded as
\begin{equation}
 n(x,t) =\overline{n}\equiv N/V,\label{165241_5Mar14}
\end{equation}
in the mixing
process, where $N$ is the total number of particles in the system and
$V$ is the volume of the system. 

From the linear irreversible thermodynamics
(see Chap. ~XI-7 in Ref.~\cite{Degroot}),
when the system is uniform in the
$y$-direction and the external field does not
exist, the heat current $J_Q$ and
the diffusion current $\bar{J}_A^m$ of the component\ $A$,
which is defined as
$\bar{J}_k^m \equiv n_k( v_k -v^m)$ where
$v^m \equiv \sum_k x_k v_k$ is the mean velocity and
$v_k$ is the velocity of the component $k$ in the $x$-direction, 
are written as
\begin{align}
 J_Q &= - \kappa \frac{\partial T}{\partial x} - n_A  T D''
 \textcolor{black}{\tilde{\mu}}^x_{AA} \frac{\partial x_A}{\partial x}, \label{172052_12Dec13}\\ 
 \bar{J}_A^m &= - n x_A x_B D' \frac{\partial T}{\partial x}
 - n D  \frac{\partial x_A}{\partial x},\label{183054_26Jun13} 
\end{align}
where $D'$ and $D$ denote the thermal diffusion coefficient and the
diffusion coefficient,  respectively,
$\textcolor{black}{\tilde{\mu}}_A$
is the chemical potential per particle of the
component $A$, and
$\textcolor{black}{\tilde{\mu}}^x_{AA} \equiv $
$( \partial  \textcolor{black}{\tilde{\mu}}_A / \partial x_A )_{T,p}$.
Equations~(\ref{172052_12Dec13}) and (\ref{183054_26Jun13}) can be derived by
taking ~the thermodynamic forces as \textcolor{black}{$-{\vec \nabla} T/T^2$} and 
\textcolor{black}{
$- \textcolor{black}{\tilde{\mu}}^x_{AA} ( {\vec \nabla} x_A ) / (x_B T ) $}.
Then, the \textcolor{black}{coefficients of the Onsager relations
\begin{align}
  J_Q &= -L_{qq} \frac{1}{T^2}\frac{\partial T}{\partial x}
  - L_{qA} \frac{\textcolor{black}{\tilde{\mu}}^x_{AA}}{x_BT} \frac{\partial x_A}{\partial x}, \label{172052_17Mar1}\\ 
 \bar{J}_A^m &= -L_{Aq} \frac{1}{T^2}\frac{\partial T}{\partial x}
  - L_{AA} \frac{\textcolor{black}{\tilde{\mu}}^x_{AA}}{x_BT} \frac{\partial x_A}{\partial x},\label{183054_17Mar1} 
\end{align}
}are written as
\textcolor{black}{
$L_{qq}=\kappa T^2$, $L_{qA}=n x_A x_B T^2D''$,
$L_{Aq}=n x_A x_B T^2D'$ and 
$L_{AA}=n  x_B TD/\textcolor{black}{\tilde{\mu}}^x_{AA}$},
 thus the Onsager reciprocal relation leads to $D''=D'$.
Additionally, if $v^m=0$ holds, the time evolution equations of $T$ and
$x_A$ written as
\begin{align}
 c_p  \frac{\partial T}{\partial t} &= \frac{\partial }{\partial x}
  \big\{ \kappa \frac{\partial T}{\partial x} + n_A  T D''
  \textcolor{black}{\tilde{\mu}}^x_{AA} \frac{\partial x_A}{\partial x}  \big\},\label{001112_6Dec13} \\
   n  \frac{\partial x_A}{\partial t} &=  \frac{\partial }{\partial x} \big\{
    n x_A x_B D' \frac{\partial T}{\partial x}  + n D  \frac{\partial
 x_A}{\partial x} \big\},\label{184739_26Jun13} 
\end{align}
which can be derived from the conservation laws of energy and mass
(see Chap.XI-7 in Ref.~\cite{Degroot}), respectively.
Here, $c_p$ is the specific heat at constant pressure per
unit volume.

The time evolution equations of $x_A$ and $T$ in the mixing process can
be derived from equations~(\ref{001112_6Dec13}) and (\ref{184739_26Jun13})
by using $\partial p/\partial x=\partial n/\partial x=0$
(in Eqs.~(\ref{165229_5Mar14}) and (\ref{165241_5Mar14})), and then can be
simplified by neglecting the second-order terms of 
$\partial T/\partial x$ and $\partial x_A/\partial x$, as
\begin{align}
 c_p \frac{\partial T}{\partial t} &= 
 l_{11}(x,t) \,\frac{\partial^2 T}{\partial x^2}
 + l_{12}(x,t) \, \frac{\partial^2 x_A}{\partial x^2},\label{225422_23Nov11}  \\
 n \frac{\partial x_A}{\partial t}  &= 
 l_{21}(x,t) \, \frac{\partial^2 T}{\partial x^2}
 + l_{22}(x,t) \, \frac{\partial^2 x_A}{\partial x^2},\label{225426_23Nov11}
\end{align}
where $l_{11} \equiv \kappa$,
$l_{12} \equiv n_A \textcolor{black}{\tilde{\mu}}^x_{AA} T D''$,
$l_{21} \equiv n x_A x_B D'$, and $l_{22} \equiv  n D$.
We note that the coefficients $l_{ij}$
depend on the position $x$ and the time $t$ through
$p,T,x_A$ and $n$.
These time evolution equations \textcolor{black}{should be} solved under the boundary
conditions
\begin{align}
 \bar{J}_A^m (0,t)=\bar{J}_A^m (L_x,t)=0,\label{182332_12Dec13}\\
 T(0,t)=T(L_x,t)=T_0,\label{085933_13Dec13}~
\end{align}
using equation~(\ref{183054_26Jun13}).
Since $\Delta t_{\text{sep}} \gg \tau_{\text{sep}}$,
the initial condition of the mixing process is written as
\begin{align}
 T(x,0)=T_0,~~
 x_A(x,0)=x_A^E(x),\label{132905_17Jun13}
\end{align}
where $x_A^E(x)$ denotes the mole fraction profile of the equilibrium
state in the end of the separating process with the external field $E$
\textcolor{black}{and we note that $t=0$ is chosen as
the beginning of the mixing process unlike Figures 3-6}.
Similarly, because of $\Delta t_{\text{mix}} \gg \tau_{\text{mix}}$,
the profiles of the mole fraction $x_A$ and the
temperature $T$ in the end of the mixing process are written as
\begin{align}
 T(x,\Delta t_{\text{mix}})=T_0,~~
 x_A(x,\Delta t_{\text{mix}})=\overline{x}_A,\label{132923_17Jun13}
\end{align}
where $\overline{x}_A\equiv N_A/N$ is the mean mole fraction in the
system. 

The cooling power (\ref{123022_11Dec13}) is expressed as
\begin{align}
 \overline{\dot{Q}}_c \equiv
 \frac{-1}{\Delta t_{\text{sep}}+\Delta t_{\text{mix}}}
 \int_0^{\Delta t_{\text{mix}}} L_y J_Q(L_x,t)dt
 ,\label{004427_16Dec13}
\end{align}
\textcolor{black}{after the steady cyclic state is established.}
By using equations~(\ref{172052_12Dec13}) and (\ref{183054_26Jun13}) with the
coefficients $l_{ij}$ and the boundary condition (\ref{182332_12Dec13}), we can obtain
\begin{equation}
    \overline{\dot{Q}}_c =
    \frac{L_y
 \int_0^{\Delta t_{\text{mix}}}
    \big( l_{11}-l_{12} \frac{l_{21}}{l_{22}} \big)
    \frac{\partial T}{\partial x}(L_x,t) dt
 }{\Delta t_{\text{sep}}+\Delta t_{\text{mix}}}
    .\label{003615_24Nov11}
\end{equation}
To obtain the expression for the COP, we write equation~(\ref{003700_24Nov11})
 as 
\begin{align}
  \epsilon = \frac{\overline{\dot{Q}}_c}{W_E/(\Delta t_{\text{sep}}+\Delta
 t_{\text{mix}})} ,\label{021440_13Dec13}
\end{align}
using the relation
  $\overline{\dot{W}}=W_E/(\Delta t_{\text{sep}}+\Delta t_{\text{mix}})$,
where $W_E$ denotes the total work done by the external field $E_x=E$ 
in the separating process. The work $W_E$ is written as
\begin{align}
   W_E = \psi_E^{\text{initial}} 
 - \psi_E^{\text{final}},\label{184813_12Dec13}
\end{align}
where $\psi_E^{\text{initial}}$ and $\psi_E^{\text{final}}$ are the
potential energies due to the electric field $E_x=E$ in the
initial and final states, respectively, of the separating process
(see Appendix \ref{sec:apdx_W_E}).
Using the profiles $x_A(x)$ and $n(x)$ of the system, the potential
$\psi_E$ is given by
\begin{align}
 \psi_E [x_A(x),n(x)] = qEL_y \int_0^{L_x} n(x) \big( 2x_A(x)-1 \big) x \,
 dx.\label{190101_12Dec13} 
\end{align}
Since, in the separating process, the initial profiles of
$x_A(x)$ and $n(x)$ are $\overline{x}_A$ and $\overline{n}$,
respectively, and the final profiles are
$x^E_A(x)$ and $n^E(x)$, where $n^E(x)$ is defined similarly to
$x^E_A(x)$ below equation~(\ref{132905_17Jun13}),
equation~(\ref{184813_12Dec13}) becomes 
\begin{align}
 W_E&=-qEL_y \int_0^{L_x} \Big\{ \delta n^E(x) \big( 2\overline{x}_A -1 \big)
  \nonumber \\  &\qquad\quad
 + 2 \overline{n} \, \delta x_A^E(x)
 + 2 \delta n^E(x) \delta x_A^E(x) \Big\}x\,dx,\label{003546_24Nov11}
\end{align}
where we defined
$\delta x_A^E(x) \equiv x_A^E(x) - \overline{x}_A$ and
$\delta n^E(x) \equiv  n^E(x)-\overline{n}$.
Therefore, by substituting equation~(\ref{003546_24Nov11}) into 
equation~(\ref{021440_13Dec13}), the expression for the COP is written as
\begin{widetext}
 \begin{equation}
 \epsilon =
 \frac{\int_0^{\Delta t_{\text{mix}}}
 \Big( -l_{11}+l_{12} \frac{l_{21}}{l_{22}} \Big)
 \frac{\partial T}{\partial x}(L_x,t) dt}{
 qE \int_0^{L_x} \Big\{
 \big( 2\overline{x}_A -1 \big) \delta n^E(x)
 + 2 \overline{n} \, \delta x_A^E(x)
 + 2 \delta n^E(x) \delta x_A^E(x)
 \Big\}x\,dx}.\label{005003_24Nov11}
 \end{equation}
\end{widetext}
\subsection{\label{sec:COPcp_2}Approximate calculation of the Cooling Power and the COP}
We make two assumptions to calculate $\overline{\dot{Q}}_c$ and
$\epsilon$ approximately.
The first assumption is that 
$\partial T/\partial x$,\,$\partial x_A/\partial x$ and 
$E$ are very small so that the coefficients
$l_{ij},c_p$ and $n$
in the time evolution equations (\ref{225422_23Nov11}) and (\ref{225426_23Nov11}) approximately
depend only on the average values over the system,
not on the time and the position.
Under this assumption, we write $l_{ij},c_p$ and $n$ as
$\overline{l}_{ij}$, $\overline{c}_p$ and 
$\overline{n}$, respectively, in the following.
Therefore, we can linearize 
equations~(\ref{225422_23Nov11}) and (\ref{225426_23Nov11}) with
the constants
$\overline{l}_{ij}$, $\overline{c}_p$ and $\overline{n}$ as
\begin{align}
 \overline{c}_p \frac{\partial T}{\partial t} (x,t) &= 
 \overline{l}_{11}\frac{\partial^2 T}{\partial x^2}  (x,t)
 + \overline{l}_{12} \frac{\partial^2 x_A}{\partial x^2} (x,t),\label{014819_24Nov11} \\
 \overline{n} \frac{\partial x_A}{\partial t} (x,t)  &= 
 \overline{l}_{21} \frac{\partial^2 T}{\partial x^2} (x,t)
 + \overline{l}_{22} \frac{\partial^2 x_A}{\partial x^2} (x,t).\label{014824_24Nov11} 
\end{align}
We can calculate the cooling power (\ref{003615_24Nov11})
by solving these time evolution equations
\textcolor{black}{(\ref{014819_24Nov11}) and (\ref{014824_24Nov11}) of the mixing process}
without using the similar
equations of the separating process, because the heat does not flow from
the cold heat bath in the separating process.
The second assumption is that the mixture can be regarded as an ideal gas
when the system is in the equilibrium state.
Using the second assumption and the equilibrium statistical mechanics,
$\delta n^E(x)$ and $\delta x_A^E(x)$ defined below equation~(\ref{003546_24Nov11})
can be calculated as
\begin{align}
 \delta n^E(x) &=
 \frac{\beta Eq}{L_y} \frac{
    N_A e^{\beta Eq (\frac{L_x}{2}-x)} + N_B e^{-\beta Eq (\frac{L_x}{2}-x)}}
    {e^{\beta Eq \frac{L_x}{2}}-e^{-\beta Eq \frac{L_x}{2}}}
     -\overline{n}
 \\
 &\simeq
 (2\overline{n}_A-\overline{n})\beta Eq \big( \frac{L_x}{2}-x
 \big),\label{135913_17Jun13}\\
 \delta x_A^E(x) &=
 \frac{ N_A e^{\beta Eq (\frac{L_x}{2}-x)} }
    {N_A e^{\beta Eq (\frac{L_x}{2}-x)} + N_B e^{-\beta Eq
   (\frac{L_x}{2}-x)}}-\overline{x}_A
 \\
 &\simeq
 2\overline{x}_A(1-\overline{x}_A)\beta Eq \big( \frac{L_x}{2}-x
 \big),\label{133438_17Jun13} 
\end{align}
where $\beta \equiv 1/k_B\overline{T}$ and
$\overline{T}=T_0$, and we expanded the equations up to
the first order of $E$. 
From the assumption of ideal gas, we can obtain
$\textcolor{black}{\tilde{\mu}}^x_{AA}=k_BT/x_A$, therefore
\begin{align}
    \overline{l}_{12}=k_B\overline{T}^2\overline{n}D''.\label{013838_16Aug01}
\end{align}
Similarly, we can obtain the relations
\begin{align}
  \overline{l}_{21} &= \overline{n}\, \overline{x}_A \overline{x}_B D',\label{013839_16Aug01}\\
  \overline{l}_{22} &=  \overline{n} D.\label{013840_16Aug01}
\end{align}
We next give the integral
$\int_0^{\Delta t_{\text{mix}}}(\partial T/\partial x)(L_x,t)\,dt$
in the expression for
the cooling power (\ref{003615_24Nov11})
By eliminating $\partial^2x_A/\partial x^2$ from
equations~(\ref{014819_24Nov11}) and (\ref{014824_24Nov11}),
we obtain
\begin{align}
 \overline{c}_p \frac{\partial T}{\partial t}=
   \overline{l}_{1}'  \frac{\partial^2 T}{\partial x^2}+
  \overline{l}_{2}' \frac{\partial x_A}{\partial t},\label{013838_13Dec13}
\end{align}
where 
\begin{align}
  \overline{l}_{1}'
  &
  \equiv \overline{l}_{11} - \overline{l}_{12} \overline{l}_{21}\big/\overline{l}_{22},
  \label{013841_16Aug01}\\
  \overline{l}_{2}'
  &
  \equiv \overline{l}_{12} \overline{n}\big/\overline{l}_{22},
    \label{013842_16Aug01}
\end{align}
are introduced for simplicity.
By integrating equation~(\ref{013838_13Dec13}) with respect to the time $t$ on
$[0,\Delta t_{\text{mix}}]$, 
we obtain
\begin{align}
 0=\overline{l}_{1}' \frac{\partial^2}{\partial x^2}
 \int_0^{\Delta t_{\text{mix}}} T(x,t) dt +
 \overline{l}_{2}' \Big( -\delta x_A^E (x)\Big).\label{021807_24Nov11}
\end{align}
The above equation can be integrated with respect to $x$ by
substituting $\delta x_A^E(x)$ of equation~(\ref{133438_17Jun13}) into
equation~(\ref{021807_24Nov11}) and using the boundary condition
(\ref{085933_13Dec13}).
Then, we obtain
\begin{align}
 &\int_0^{\Delta t_{\text{mix}}} T(x,t) dt=
 \frac{2\overline{l}_{2}' \overline{x}_A(1-\overline{x}_A) \beta Eq}{\overline{l}_{1}'}
 \nonumber \\ &\qquad\qquad \times
 \Big( -\frac{x^3}{6} +\frac{L_x}{4}x^2 -\frac{L_x^2}{12}x \Big) + T_0
 \Delta t_{\text{mix}}.
\end{align}
Therefore, the cooling power is written as
\begin{align}
 \overline{\dot{Q}}_c =
 \frac{-k_B \overline{T}^2 \overline{n}D''\overline{x}_A(1-\overline{x}_A) \beta EqL_yL_x^2}{
 6D(\Delta t_{\text{sep}}+\Delta t_{\text{mix}})}.
 \label{142803_17Jun13}
\end{align} 
By substituting equations~(\ref{135913_17Jun13}) and (\ref{133438_17Jun13})
into equation~(\ref{003546_24Nov11}), and expanding 
up to the second order of $E$, the work $W_E$ becomes
\begin{align}
 W_E= \frac{L_y L_x^3 \beta (qE)^2 \overline{n}}{12}.\label{142751_17Jun13}
\end{align}
Consequently, the COP in equation~(\ref{021440_13Dec13}) is written as
\begin{align}
 \epsilon 
 = \frac{-2k_B \overline{T}^2 D''\overline{x}_A(1-\overline{x}_A)  }{
 L_xqED}.\label{134139_1Jul13}
\end{align}

\subsection{\label{sec:md2}Numerical confirmation}
\begin{figure}
\resizebox{0.48\textwidth}{!}{%
  \includegraphics{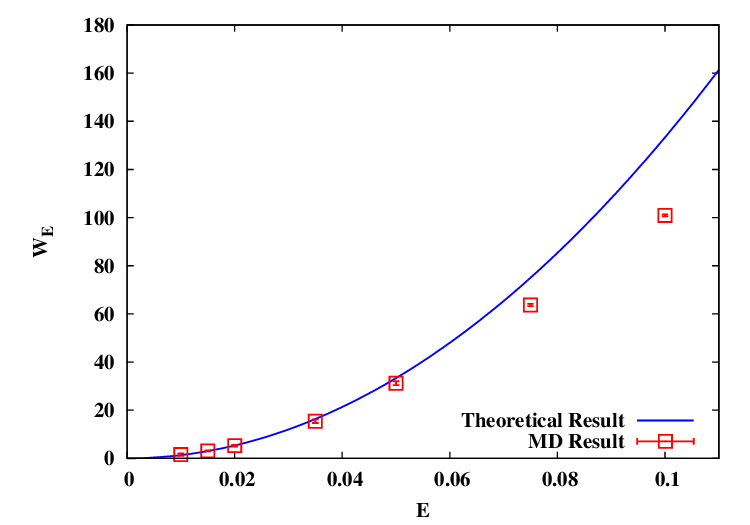}
}
\caption{\label{W-E_Nbin5_20130912_1213}
  Comparison between the theoretical result (\ref{142751_17Jun13})
 and the MD result of the work $W_E$ done by the external field $E$ to
 the system. The MD data were averaged over 2000-7500 cycles.
 }
\end{figure}
To compare the theoretical results (\ref{142803_17Jun13}) and
(\ref{134139_1Jul13}) with
the MD results,
the transport coefficients such as $D$ and $D''$ need
to be determined.
It is convenient to introduce the thermal diffusion ratio $k_T$ defined
as 
\begin{align}
 k_T \equiv \overline{T} \overline{x}_A \overline{x}_B \frac{D'}{D},
\end{align}
because our main results (\ref{142803_17Jun13}) and
(\ref{134139_1Jul13}) can be rewritten with only $k_T$ instead of $D$
and $D''(=D')$ as 
\begin{align}
 \overline{\dot{Q}}_c 
 &=\frac{-k_T N EqL_x}
 {6(\Delta t_{\text{sep}}+\Delta t_{\text{mix}})},\label{164241_14Dec13}\\
 \epsilon 
 &= \frac{-2k_B k_T\overline{T}}{L_xqE},\label{164248_14Dec13}
\end{align}
respectively.

$k_T$ can approximately be calculated from the
Chapman-Enskog theory (see Appendix \ref{sec:apdx_Chap}).
We numerically calculated the two-dimensional expression for $k_T$ in
the first order approximation as
\begin{align}
 k_T \simeq -0.13657,\label{164104_14Dec13}
\end{align}
using the parameters $m_A=1,m_B=10,\overline{T}=1$, 
$\overline{x}_A=\overline{x}_B=0.5$, and $Y=10^5$ of the Herzian
potential.

In the MD simulations in this section, the numbers
of particles are changed to $N_A=N_B=50$ so that the number-density of
the particles in the system becomes 
adequately dilute, which is assumed in the Chapman-Enskog theory.
The calculation of equation~(\ref{164104_14Dec13}) is also valid for these
new parameters.
$\Delta t_{\text{sep}}$ and $\Delta t_{\text{mix}}$ are fixed
to 10000 and 5000, respectively, so that the assumption of $\Delta t_{\text{sep}} \gg \tau_{\text{sep}}$
and $\Delta t_{\text{mix}} \gg \tau_{\text{mix}}$ is satisfied.
All other parameters such as the system size are identical with Section
\ref{sec:model}.

Figure~\ref{W-E_Nbin5_20130912_1213} shows the numerical result of
the work $W_E$
done by the external field $E_x=E$ as varying $E$, together
 with the theoretical result (\ref{142751_17Jun13}).
From Figure~\ref{W-E_Nbin5_20130912_1213}, we can see that the MD
data deviate from the theoretical curve when $0.07 \lesssim E$. This
result implies that the assumption of small $E$ in our theory is not
satisfied when $0.07 \lesssim E$ and the consequent results of the theory
may not be accurate. 
This is because the number-density in some parts of the mixture
becomes high and the mixture deviates from ideal gas when the external
 field $E$ is large. 
\begin{figure}[b]
\resizebox{0.48\textwidth}{!}{%
  \includegraphics{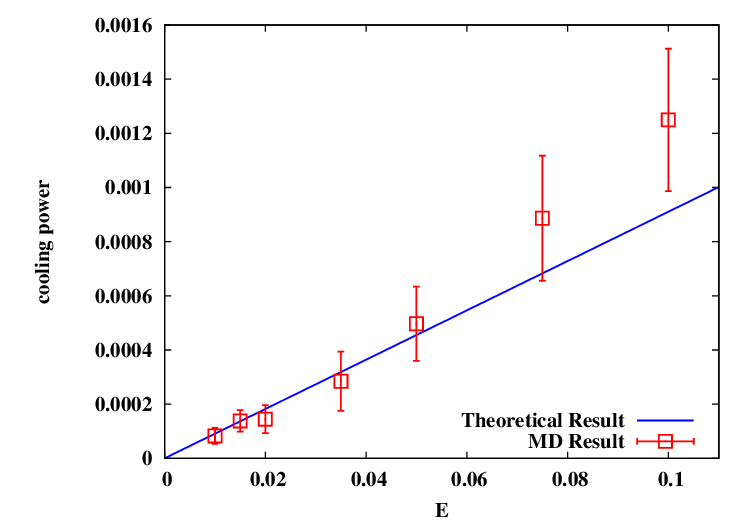}
}
\resizebox{0.48\textwidth}{!}{%
  \includegraphics{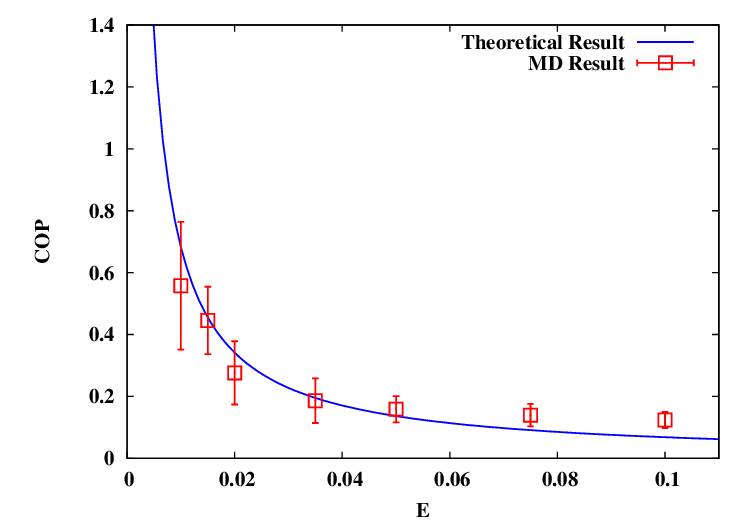}
}
\caption{\label{QcCOP-E_Nbin5_20130912_1213}
  Comparison between the theoretical results (\ref{164241_14Dec13})
 and (\ref{164248_14Dec13}) 
 and the MD results of the cooling power $\overline{\dot{Q}}_c$ and the
 COP $\epsilon$, respectively.
 The MD data were averaged over 2000-7500 cycles.
 }
\end{figure}

The theoretical results of the cooling power $\overline{\dot{Q}}_c$ and
the COP $\epsilon$ in   
equations~(\ref{164241_14Dec13}) and (\ref{164248_14Dec13}) using the value of 
equation~(\ref{164104_14Dec13}) are compared with the MD data in
Figure~\ref{QcCOP-E_Nbin5_20130912_1213}.
We can confirm a good agreement between the theory and the MD data in
the region $E \lesssim 0.05$, but a small discrepancy in the region
$0.07 \lesssim E$ where the assumption of small $E$ may not be satisfied.
Therefore, we conclude that the validity of our theoretical analysis
of the heat pump model is verified in the case of $T_h=T_c$.

\subsection{\label{sec:COPcp_3}The case of $T_h \neq T_c$}
Finally, we show that the theoretical analysis in
Sections \ref{sec:COPcp_1} and \ref{sec:COPcp_2} can be
generalized to the case of $T_h \neq T_c$.
We consider the case that the temperature difference of the heat baths
$\delta T \equiv T_h-T_c$ is very small,
and $\Delta t_{\text{sep}} \gg \tau_{\text{sep}}$ and 
 $\Delta t_{\text{mix}} \gg \tau_{\text{mix}}$ are satisfied.
The time evolution equations (\ref{225422_23Nov11}) and
(\ref{225426_23Nov11}) in the mixing process hold even in this
case, and we assume that the linear approximation in
equations~(\ref{014819_24Nov11}) and (\ref{014824_24Nov11}) is
also valid.
The boundary condition (\ref{182332_12Dec13}) is unchanged, but 
equation~(\ref{085933_13Dec13}) should be changed to 
\begin{align}
 T(0,t)=T_{\text{h}},\quad T(L_x,t)=T_{\text{c}}.\label{091753_13Dec13}
\end{align}
The initial conditions of $T(x,t)$ and $x_A(x,t)$ in the mixing
process are
\begin{align}
 T(x,0)=T_{\text{h}},~~
 x_A(x,0)=x_A^E(x),\label{132906_17Jun13} 
\end{align}
respectively.

The profiles $T(x,t)$ and $x_A(x,t)$ in the end of
the mixing process in equation~(\ref{132923_17Jun13}) become
\begin{align}
 T(x,\Delta t_{\text{mix}})=T^{\delta T}(x),~~
 x_A(x,\Delta t_{\text{mix}})=x_A^{\delta T}(x),\label{132924_17Jun13}
\end{align}
where $T^{\delta T}(x)$ and $x_A^{\delta T}(x)$ denote
the temperature and the mole fraction profiles, respectively, of 
the steady heat-conduction state in the mixing process when the temperature
difference between the heat baths $\delta T$ exists.
In the steady state of the mixing process,
the temperature profile $T^{\delta T}(x)$ is written as
\begin{equation}
 T^{\delta T}(x)=-\frac{\delta T}{L_x}x+T_{\text{h}},\label{093518_13Dec13}
\end{equation}
which can be derived from the time evolution equations 
(\ref{014819_24Nov11}) and (\ref{014824_24Nov11}) and the 
boundary condition (\ref{091753_13Dec13}).
To determine the mole fraction profile $x_A^{\delta T}(x)$, we need an
additional assumption, besides the assumptions in
Sections~\ref{sec:COPcp_1} and \ref{sec:COPcp_2},
that each local subsystems of the mixture can be regarded as equilibrium
ideal gas in that subsystem when the system is in the steady state of the mixing process.
Using the equation of state of ideal gas, we can write
\begin{align}
 n^{\delta T}(x)=\frac{p(x)}{k_BT^{\delta T}(x)}
 \simeq \frac{p(x)}{k_BT_{\text{h}}} \bigg(1 + \frac{\delta T}{T_{\text{h}}L_x}x\bigg),\label{095833_8Jul13}
\end{align}
where $p(x)$ is the pressure profile, $n^{\delta T}(x)$ denotes the
 number-density profile of the steady state in the mixing process,
and we neglected the terms $\mathcal{O}(\delta T^2)$. 
Using the assumption of the mechanical equilibrium state
${\vec \nabla}p=0$ in Section~\ref{sec:COPcp_1} and
the relation $N = \int_{0}^{L_x} L_y n^{\delta T}(x) dx$,
equation~(\ref{095833_8Jul13}) can be rewritten as
\begin{align}
 n^{\delta T}(x) = \overline{n}+
 \frac{\overline{n} \delta T}{\overline{T}L_x}
 \bigg(x - \frac{L_x}{2}\bigg),\label{101426_10Jul13}    
\end{align}
where $\overline{T}\equiv (T_h+T_c)/2$.
When the system is in the steady state, the linear relation
(\ref{183054_26Jun13}) becomes
\begin{equation}
 0= -  \overline{l}_{21} \frac{\partial T^{\delta T}}{\partial x} (x)
  - \overline{l}_{22} \frac{\partial x_A^{\delta T}}{\partial x} (x).\label{110340_13Dec13}
\end{equation}
Thus, from equation~(\ref{093518_13Dec13}) and the relation
\begin{equation}
 N_A= \int_{0}^{L_x} L_y n^{\delta T}(x)x_A^{\delta T}(x)dx, 
\end{equation}
the mole fraction profile $x_A^{\delta T}(x)$ is written as
\begin{align}
 x_A^{\delta T}(x) = \frac{\overline{x}_A \overline{x}_BD'\delta T}{DL_x}
 \Big(x- \frac{L_x}{2}\Big)
 +\overline{x}_A,\label{094943_11Jul13}
\end{align}
where we use
equations~(\ref{013839_16Aug01}) and (\ref{013840_16Aug01}).

The cooling power $\overline{\dot{Q}}_c$ can be calculated in the same
way as in Sections~\ref{sec:COPcp_1} and \ref{sec:COPcp_2},
 but the condition $\delta T\neq 0$ changes
equation~(\ref{021807_24Nov11}) to 
\begin{align}
 &\overline{c}_p \big(T^{\delta T}(x)-T_{\text{h}}\big)
 =\overline{l}_{1}' \frac{\partial^2}{\partial x^2} 
 \int_0^{\Delta t_{\text{mix}}}  T(x,t) dt
 \nonumber \\ &\qquad\qquad\qquad\qquad\qquad
 + \overline{l}_{2}' \big( \delta x_A^{\delta T}(x)
 -\delta x_A^E (x)\big),\label{021806_24Nov11}
\end{align}
where
$\delta x_A^{\delta T}(x)\equiv x_A^{\delta T}(x) -\overline{x}_A$.
Since the expression (\ref{003615_24Nov11}) is valid even in the
present case, the cooling power is obtained as
\begin{widetext}
 \begin{equation}
  \overline{\dot{Q}}_c
   =\frac{-L_y}{6(\Delta t_{\text{sep}}+\Delta t_{\text{mix}})}
   \Big\{2 \overline{c}_p \delta T L_x
   +\frac{\overline{l}_{2}' \overline{x}_A \overline{x}_B D''\delta TL_x}{2D}
   +\overline{l}_{2}' \overline{x}_A \overline{x}_B
   \beta \Big(1-\frac{\delta T}{2 \overline{T} }\Big) EqL_x^2
   + \frac{6 \overline{l}_{1}'\delta T \Delta t_{\text{mix}}}{L_x}  \Big\},\label{122942_13Dec13}
 \end{equation}
\end{widetext}
by ~~solving ~~the ~~differential ~~equation~~(\ref{021806_24Nov11}) ~~for
$\int_0^{\Delta t_{\text{mix}}} T(x,t) dt$.
We note that $\overline{l}_{1}'$ in equation~(\ref{013841_16Aug01}) is positive
since $L_{qA}^2<L_{qq}L_{AA}$
 \cite{Prigogine}.

 In the case of $\delta T\neq 0$, the expression for $W_E$ in
equation~(\ref{003546_24Nov11}) becomes
 \begin{align}
  W_E&
  =\psi_E [\overline{x}_A+ \delta x_A^{\delta T}(x),
  \overline{n}+ \delta n^{\delta T}(x)]
  \nonumber \\&\qquad\qquad\qquad\quad
    - \psi_E [\overline{x}_A+ \delta x_A^E(x),\overline{n}+ \delta
  n^E(x)]
  \\
  &=qEL_y \int_0^{L_x} \Big\{
  \big(\delta n^{\delta T} (x)-\delta n^E(x) \big) \big( 2\overline{x}_A -1 \big)
  \nonumber \\ &
  + 2 \overline{n} \, \big(\delta x_A^{\delta T} (x)-\delta x_A^E(x)
  \big)
  +\mathcal{O}(\delta T^2)
  +\mathcal{O}(E^2)
  \Big\}x\,dx,\label{003547_24Nov11} 
 \end{align}
where
$\delta n^{\delta T}(x)\equiv n^{\delta T}(x) -\overline{n}$.
 By substituting equations~(\ref{135913_17Jun13}), (\ref{133438_17Jun13}),
  (\ref{101426_10Jul13}) and (\ref{094943_11Jul13}) into
  equation~(\ref{003547_24Nov11}), we can obtain 
 \begin{align}
  W_E&\simeq \frac{\overline{n} qEL_yL_x^2}{12} \bigg\{
     \frac{(2\overline{x}_A-1 +2\overline{x}_A \overline{x}_B\overline{T}D'/D)\delta T}
     {\overline{T}}
  \nonumber\label{232644_3Mar14} \\ &\qquad\qquad\qquad\qquad\qquad
     + \beta
   \Big(1-\frac{\delta T}{2 \overline{T} }\Big) EqL_x
     \bigg\}.
 \end{align}
Substituting equations~(\ref{122942_13Dec13}) and (\ref{232644_3Mar14}) into
equation~(\ref{021440_13Dec13}), we finally obtain the COP as 
\begin{widetext}
    \begin{align}
   \epsilon
   &=\frac{-2}{\overline{n} qEL_x^2} \,
     \frac{
2 \overline{c}_p \delta T L_x
   +\frac{\overline{l}_{2}' \overline{x}_A \overline{x}_B D''\delta TL_x}{2D}
   +\overline{l}_{2}' \overline{x}_A \overline{x}_B
   \beta \Big(1-\frac{\delta T}{2 \overline{T} }\Big) EqL_x^2
   + \frac{6 \overline{l}_{1}'\delta T \Delta t_{\text{mix}}}{L_x}
     }{
     \frac{(2\overline{x}_A-1 +2\overline{x}_A \overline{x}_B\overline{T}D'/D)\delta T}
     {\overline{T}}+ \beta \Big(1-\frac{\delta T}{2 \overline{T} }\Big) EqL_x
     }
     .\label{134245_13Dec13}
  \end{align}
\end{widetext}
Since $\overline{l}_{1}'>0$, equation~(\ref{134245_13Dec13}) means that the
longer $\Delta t_{\text{mix}}$ is, the lower $\epsilon$ becomes because the
heat begins to flow in the reverse direction after a temperature
gradient is established due to the temperature difference of the heat
baths.  

\begin{figure}
\resizebox{0.48\textwidth}{!}{%
  \includegraphics{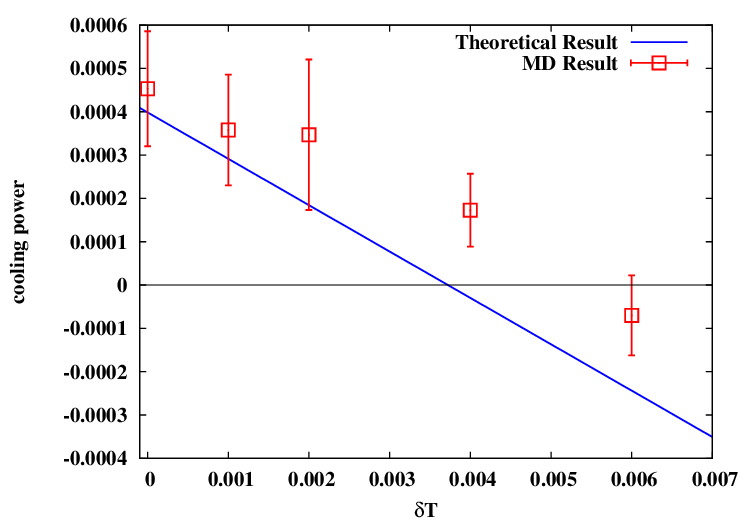}
}
\resizebox{0.48\textwidth}{!}{%
  \includegraphics{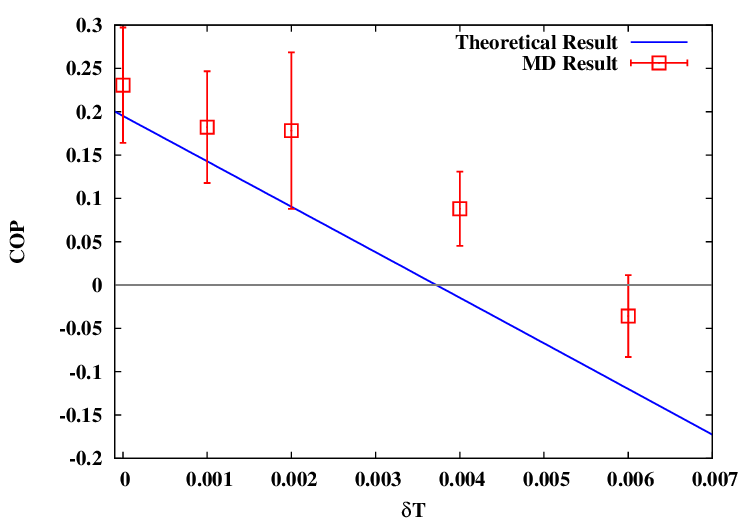}
}
\caption{\label{2dmd_COPcp-DelT_2}
 The temperature difference $\delta T=T_h-T_c$ dependence of the
 cooling power and the COP, with the process durations
 $\Delta t_{\text{sep}}=5000,\Delta t_{\text{mix}}=3000$,
 and the temperatures $T_h=1+\delta T/2$ and $T_c=1-\delta T/2$.
 The MD data were averaged over 6000 cycles.
 }
\end{figure}

Finally, we compare the theoretical results in this section with
  the MD results.
  By using equations (\ref{013841_16Aug01}) and (\ref{013842_16Aug01}), equations (\ref{122942_13Dec13}) and (\ref{134245_13Dec13}) can be rewritten as
\begin{widetext}
    \begin{align}
\overline{\dot{Q}}_c&=\frac{-L_y}{6(\Delta t_{\text{sep}}+\Delta t_{\text{mix}})}
   \Bigg[\Big\{2 \overline{c}_p  L_x
   +
   \frac{k_B \overline{n} k_T^2 L_x }{2\overline{x}_A \overline{x}_B}
  -\frac{\overline{n} k_TEqL_x^2}{2 \overline{T} } 
   + \frac{6 \lambda \Delta t_{\text{mix}}}{L_x}
  \Big\}\delta T
  + \overline{n} k_T EqL_x^2 \Bigg],\label{231359_10Mar14}\\
  \epsilon &=\frac{-2}{\overline{n} qEL_x^2} \,
  \frac{
  \Big\{2 \overline{c}_p  L_x
   +
   \frac{k_B \overline{n} k_T^2 L_x }{2\overline{x}_A \overline{x}_B}
  -\frac{\overline{n} k_TEqL_x^2}{2 \overline{T} } 
   + \frac{6 \lambda \Delta t_{\text{mix}}}{L_x}
  \Big\}\delta T
  + \overline{n} k_T EqL_x^2
  }{
  \Big\{ 2\overline{x}_A-1 + 2k_T - \frac{\beta EqL_x}{2}
  \Big\}\frac{\delta T}{\overline{T}} +\beta EqL_x},\label{231433_10Mar14}
\end{align}
\end{widetext}
respectively, 
where we introduced the coefficient $\lambda$ defined as
\begin{align}
  \lambda = \kappa - \overline{n}k_B\overline{T}^2
  \textcolor{black}{\overline{x}_A\overline{x}_B} \frac{D'^2}{D} ,\label{231433_16Aug21}
\end{align}
 which can be calculated in the first order approximation as
\begin{align}
 \lambda \simeq 0.419877,\label{164104_16Aug21}
\end{align}
by using its microscopic expression (\ref{203590_16Aug15})
with the same parameters as used in equation~(\ref{164104_14Dec13}).
Figure~\ref{2dmd_COPcp-DelT_2} shows
the MD results of
the cooling power and
  the COP as varying the temperature difference $\delta T$, together with the
  theoretical results (\ref{231359_10Mar14}) and (\ref{231433_10Mar14}) using
  equation (\ref{164104_16Aug21}) and $\overline{c}_p=2k_B\overline{n}$,
  which is the two-dimensional ideal-gas value.
  In the MD simulation in Figure~\ref{2dmd_COPcp-DelT_2}, the external field $E$ was changed
  to $E=0.035$ from $E=0.1$ of Figure~\ref{2dmd_COPcp-DelT}
  because our theory is valid when $E$ is sufficiently small.
From this figure, we can see that the theory agrees with the MD data
in the region of small $\delta T$, which shows that our theory is valid
not only in the case of $\delta T=0$ in Figure~\ref{QcCOP-E_Nbin5_20130912_1213},
but also in the case of $\delta T \ne 0$.

\section{\label{sec:summary}Summary}
 We proposed a heat pump model utilizing the Dufour effect
 and studied it by using the MD simulation and the linear irreversible
 thermodynamics. 
 This model consists of the separating process in which the
 mixture is separated by the external electric field, and the mixing
 process in which the Dufour effect occurs.
 Using the MD simulation, we calculated the cooling power and the COP of the model
 as in Figure~\ref{2dmd_COPcp-DelT},
 and numerically confirmed its usefulness as a heat pump.
 Next, we theoretically calculated the cooling power and the COP
 as equations~(\ref{164241_14Dec13}) and (\ref{164248_14Dec13}),
 especially in the simplest case of $T_h=T_c$,
 $\Delta t_{\text{sep}} \gg \tau_{\text{sep}}$ and
 $\Delta t_{\text{mix}} \gg \tau_{\text{mix}}$,
 and we confirmed a good agreement between the theoretical and MD results.
 Furthermore,
we showed that our theory is generalized to the case of $T_h \ne T_c$
 and is valid also in that case.
 
 Finally, we discuss some remaining problems. First, 
 we can find that the COP is only about 0.2\% of the Carnot COP at most from
 Figure~\ref{2dmd_COPcp-DelT}, but
 we have not yet found the conditions to obtain a heat pump
 model with much higher COP.
To know the best performance of our model, we will need
more thorough search on the parameter space of our model,
though our search in the present study was limited to where our theoretical
assumptions are probable.
 Second, it is difficult to realize our model experimentally since the
 Coulomb interaction between particles is ignored.
 To overcome this problems, our model should be generalized to consider the Coulomb interaction, for example by using MHD \cite{Moreau,THayat,AJChamkha,MNawaz}.
 We consider that experiments of our model become more realizable by
 removing the electric charges of particles and replacing the electric
 force with the gravity or inertial force such as centrifugal force
 \textcolor{black}{
   \cite{Whitley,KCohen,Kemp}.
   In this replacement, the components of a gas mixture can be separated by the pressure gradient created by the gravity or the centrifugal force}
   \footnote{\textcolor{black}{This mechanism of the separation is sometimes called the barodiffusion effect\cite{Sharma2010,Landau}.}}\textcolor{black}{.
   If a centrifuge is used, the separating process and the mixing process can be caused by accelerating and decelerating the angular velocity of the centrifuge, respectively, when
   different masses are given to the components of the gas mixture.}
 Though our model in this paper may be merely a toy model, we expect that
our work will trigger more realistic applications of the Dufour effect.

\begin{acknowledgement}
   The authors would like to thank K. Nemoto, T. Nogawa, Y. Tami, and Y. Izumida for
  fruitful discussions.
\end{acknowledgement}

\section*{Author contribution statement}
M.H. mainly contributed to all the contents of this study
including the preparation of the manuscript. K.O. supervised this study and is also responsible for all the contents.

\appendix
\numberwithin{equation}{section}

\setcounter{equation}{0}
\section{\label{sec:apdx_Chap} Two dimensional expressions for the thermal-diffusion ratio $k_T$  and the coefficient $\lambda$}
The three-dimensional microscopic expression for $k_T$ of a
binary mixture is obtained in reference~\cite{SChapman} by
approximately solving
 the subdivided Boltzmann equations by the Enskog method
(see Sect.~8 in Ref.~\cite{SChapman}).
From the similar derivation to the three-dimensional expression,
the two-dimensional expression in the first-order approximation denoted
by $[k_T]_1$ is proved to be written as
\begin{align}
  &[k_T]_1=2\Big\{x_AM_A^{-\frac{1}{2}} (a_{-1-1}a_{01}-a_{0-1}a_{1-1})
 \notag \\&\qquad\qquad\qquad
 +x_BM_B^{-\frac{1}{2}} (a_{0-1}a_{11}-a_{01}a_{1-1})
 \Big\}
 \notag \\&\qquad\qquad\qquad\qquad\qquad\qquad
 \Big/
 (a_{-1-1}a_{11}-a_{1-1}^2),\label{205257_14Dec13}
\end{align}
where $M_A\equiv m_A/(m_A+m_B)$,  $M_B \equiv m_B/(m_A+m_B)$,
and the matrix elements $a_{11}$, $a_{1-1}$, $a_{-1-1}$,
$a_{01}$ and $a_{0-1}$ in 
equation~(\ref{205257_14Dec13}) are given by
\begin{align}
 &a_{11}=x_A^2 \hat{\Omega}_{1}^{(2)}(2)
 +2x_Ax_B  \Big\{(6M_A^2M_B+4M_B^3)\hat{\Omega}_{12}^{(1)}(1)\notag\\
 &\qquad\qquad\qquad\qquad\qquad
 -4M_B^3 \hat{\Omega}_{12}^{(1)}(2) 
 +M_B^3 \hat{\Omega}_{12}^{(1)}(3)
 \notag\\ &\qquad\qquad\qquad\qquad\qquad\qquad
 +2M_AM_B^2 \hat{\Omega}_{12}^{(2)}(2)
 \Big\},
\end{align}
\begin{align}
 &a_{-1-1}=2x_Ax_B
   \Big\{(6M_B^2M_A+4M_A^3)\hat{\Omega}_{12}^{(1)}(1)
 \notag\\ &\qquad\qquad\quad
 -4M_A^3 \hat{\Omega}_{12}^{(1)}(2) 
 +M_A^3 \hat{\Omega}_{12}^{(1)}(3)
 \notag\\ &\qquad\qquad\qquad
 +2M_BM_A^2 \hat{\Omega}_{12}^{(2)}(2)
 \Big\}+ x_B^2
 \hat{\Omega}_{2}^{(2)}(2),\\
 &a_{1-1}=2x_Ax_B
   M_A^{\frac{3}{2}}M_B^{\frac{3}{2}}
 \Big\{
 - \hat{\Omega}_{12}^{(1)}(3) +4\hat{\Omega}_{12}^{(1)}(2)
 \notag\\ &\qquad\qquad\qquad\qquad\qquad
 - 10\hat{\Omega}_{12}^{(1)}(1) +2\hat{\Omega}_{12}^{(2)}(2) \Big\},\\
 &a_{01}=2x_Ax_B
 M_A^{\frac{1}{2}}\Big(2M_B^2 \hat{\Omega}_{12}^{(1)}(1)
 - M_B^2 \hat{\Omega}_{12}^{(1)}(2) \Big),\\
 &a_{0-1}=-x_Ax_B
  2M_B^{\frac{1}{2}}\Big(2M_A^2 \hat{\Omega}_{12}^{(1)}(1)
 - M_A^2 \hat{\Omega}_{12}^{(1)}(2) \Big),
\end{align}
respectively.
Here, 
$\hat{\Omega}_{12}^{(l)}(r),\hat{\Omega}_{1}^{(l)}(r)$ and $\hat{\Omega}_{2}^{(l)}(r)$
$(l,r=1,2,\cdots)$
are defined as
\begin{align}
   \hat{\Omega}_{12}^{(l)}(r) &= \frac{1}{2} \sigma
   \bigg( \frac{2 k_BT}{m_0M_AM_B} \bigg)^{\frac{1}{2}}
   \hat{\mathcal{W}}^{(l)}(r),\label{203538_12Dec13}\\
   \hat{\Omega}_{1}^{(l)}(r)&= \frac{1}{2} \sigma
   \bigg( \frac{ k_BT}{m_A} \bigg)^{\frac{1}{2}}
   \hat{\mathcal{W}}^{(l)}(r),\\
   \hat{\Omega}_{2}^{(l)}(r)&= \frac{1}{2} \sigma
   \bigg( \frac{ k_BT}{m_B} \bigg)^{\frac{1}{2}}
   \hat{\mathcal{W}}^{(l)}(r),\label{203527_12Dec13}
\end{align}
respectively,
where $m_0\equiv m_A+m_B$, $\sigma$ is the diameter of the particles,
and $\hat{\mathcal{W}}^{(l)}(r)$ are the non-dimensional values defined as
\begin{align}
  \hat{\mathcal{W}}^{(l)}(r)\equiv 2 \int_0^{\infty}
  \bigg\{ \int_0^1  e^{- g^2}
  g^{2r+1}(1-\cos^l\chi)  \,d\Big(\frac{b}{\sigma}\Big)
  \bigg\} d(g^2).\label{202155_12Dec13}
\end{align}
The parameter $\chi$ in equation~(\ref{202155_12Dec13}) is the scattering angle
between the particles with interaction potential
$U^{\text{int}}(r)$ and is a function of the scattering parameters
$g$ and $b$ written as 
\begin{align}
  \chi(g,b)=\pi -2 \int_{R}^{\infty} \bigg\{ \frac{r^4}{b^2}
 \bigg(1-\frac{U^{\text{int}}(r)}{k_BT g^2}\bigg)-r^2\bigg\}^{-\frac{1}{2}} dr,
\end{align}
where $R$ is the solution to
\begin{align}
 1-\frac{U^{\text{int}}(R)}{k_BT g^2}-\frac{b^2}{R^2}=0.
\end{align}
Using the Herzian potential in equation~(\ref{163720_14Dec13}) as
$U^{\text{int}}(r)$ above, we can finally obtain
equation~(\ref{164104_14Dec13}) as the first order approximation of $k_T$.

In the same way, the two-dimensional expression for $\lambda$
  in the first-order approximation denoted by $[\lambda]_1$ can be obtained as
  \begin{align}
    &[\lambda]_1=4k_B^2T
    \Big\{x_A^2 m_A^{-1} a_{-1-1}
      -2x_Ax_B (m_Am_B)^{-\frac{1}{2}} a_{1-1}
 \notag \\&\qquad\qquad\qquad
      +x_B^2m_B^{-1} a_{11}
 \Big\}
 \Big/
 (a_{-1-1}a_{11}-a_{1-1}^2),\label{203590_16Aug15}
\end{align}
using the similar derivation to the three-dimensional expression in reference~\cite{SChapman}.

\setcounter{equation}{0}
\section{\label{sec:apdx_W_E}
  Derivation of  equation~(\ref{184813_12Dec13})}
  To derive equation~(\ref{184813_12Dec13}), we calculate the work $W$ done to the system
  during one cycle consisting of the separating and mixing processes, which is written as
\begin{align}
  W=\sum_{i=1}^N \int d{\vec r}_{i} \cdot \Bigg\{
  q_i {\vec E} + \sum_{j(\neq i)=1}^{N}
  \big(- {\vec \nabla}_i U_{ij}^{\text{int}}\big)
  + {\vec F}_i^{\text{bath}} \Bigg\}
  , \label{B1}
\end{align}
where $U_{ij}^{\text{int}}\equiv U^{\text{int}}(|{\vec r}_i-{\vec r}_j|)$
is the interaction potential (specifically Eq. (3)) between the $i$th and $j$th particles,
${\vec F}_i^{\text{bath}}$ is the force on the $i$th particle
from the heat baths,
and the integral $\int d{\vec r}_{i}$ is evaluated along the trajectory
of the $i$th particle for one cycle of the system.
From the first term of equation~(\ref{B1}), we obtain
\begin{align}
  \sum_i \int d{\vec r}_i \cdot q_i {\vec E}
  =\psi_E^{\text{initial}}-\psi_E^{\text{final}} , \label{B2}
\end{align}
where $\psi_E$ is defined below equation (\ref{184813_12Dec13})
and we note that the mixing process does not contribute the above
equation because the electric field vanishes.
The second term of equation~(\ref{B1}) can be written as
\begin{align}
  \sum_i \int d{\vec r}_i \cdot \sum_{j(\neq i)=1}^{N}
  \big(- {\vec \nabla}_i U_{ij}^{\text{int}}\big)
 =-\int d U^{\text{int}}, 
 \label{B3}
\end{align}
where we defined $U^{\text{int}}$ as
\begin{align}
  U^{\text{int}} \equiv \frac{1}{2} \sum_{i=1}^{N}
  \sum_{j(\neq i)=1}^{N}   U_{ij}^{\text{int}}
   \label{B4}
\end{align}
Since the integral is evaluated for one cycle,
equation~(\ref{B3}) represents a change of the total interparticle potential between the
beginning and the ending of a cycle.
Therefore the second term of equation~(\ref{B1})
should macroscopically be zero as long as the system is cyclic.
Finally, the third term of equation~(\ref{B1}) can be written as
\begin{align}
  \sum_i \int d{\vec r}_i \cdot {\vec F}_i^{\text{bath}}
  = -Q_h + Q_c,
   \label{B5}
\end{align}
from the definitions of $Q_h$ and $Q_c$.

Using equations~(\ref{B2}), (\ref{B3}) and (\ref{B5}), we obtain
\begin{align}
  W=\psi_E^{\text{initial}}-\psi_E^{\text{final}} -Q_h + Q_c. \label{B6}
\end{align}
Because the denominator of the COP should be the work done by the external field except for the heat baths, $W_E$ in equation~(\ref{021440_13Dec13})
can be written as equation~(\ref{184813_12Dec13}).

% \section{Section title}
% \label{sec:1}
% and %\cite{RefJ}
% \subsection{Subsection title}
% \label{sec:2}
% as required. Don't forget to give each section
% and subsection a unique label (see Sect.~\ref{sec:1}).
% %
% % For one-column wide figures use
% \begin{figure}
% % Use the relevant command for your figure-insertion program
% % to insert the figure file.
% % For example, with the option graphics use
% \resizebox{0.75\textwidth}{!}{%
%   \includegraphics{leer.eps}
% }
% % If not, use
% %\vspace{5cm}       % Give the correct figure height in cm
% \caption{Please write your figure caption here}
% \label{fig:1}       % Give a unique label
% \end{figure}
% %
% % For two-column wide figures use
% \begin{figure*}
% % Use the relevant command for your figure-insertion program
% % to insert the figure file. See example above.
% % If not, use
% \vspace*{5cm}       % Give the correct figure height in cm
% \caption{Please write your figure caption here}
% \label{fig:2}       % Give a unique label
% \end{figure*}
% %
% % For tables use
% \begin{table}
% \caption{Please write your table caption here}
% \label{tab:1}       % Give a unique label
% % For LaTeX tables use
% \begin{tabular}{lll}
% \hline\noalign{\smallskip}
% first & second & third  \\
% \noalign{\smallskip}\hline\noalign{\smallskip}
% number & number & number \\
% number & number & number \\
% \noalign{\smallskip}\hline
% \end{tabular}
% Or use
%\vspace*{5cm}  % with the correct table height
%\end{table}
%
% BibTeX users please use
%\bibliographystyle{}
%\bibliography{}

\end{document}